\documentclass[12pt]{iopart}

\usepackage{iopams}  
\usepackage{braket}
\usepackage{stmaryrd}
\usepackage{graphicx,xcolor}
\usepackage{tikz}
\usetikzlibrary{calc}
\newcommand{\sbt}{\,\begin{picture}(-1,1)(-1,-3)\circle*{2}\end{picture}\ }
\newcommand{\deq}{\stackrel{\rm{\tiny def}}{=}}
\newcommand{\xrightarrowmd}{\stackrel{\rm mod\ 2}{\longrightarrow}}
\newcommand{\xrightarrowi}{\stackrel{{\ i\ }}{\longrightarrow}}
\newcommand{\xrightarrowpi}{\stackrel{\pi}{\longrightarrow}}
\begin{document}


\title[
Center-Preserving Automorphisms of Finite Heisenberg Group over \(\mathbb Z_N\)
]{
Center-Preserving Automorphisms of Finite Heisenberg Group over \(\mathbb Z_N\)
}


\author{Takaaki Hashimoto, Minoru Horibe and Akihisa Hayashi}

\address{Department of Applied Physics, University of Fukui, Fukui Bunkyo 3-9-1, 910-8507, Japan}
\ead{hasimoto@u-fukui.ac.jp}
\vspace{10pt}
\begin{indented}
\item[]July 2023
\end{indented}

\begin{abstract}
We investigate the group structure of center-preserving automorphisms of the finite Heisenberg group over
\(\mathbb Z_N\) with a \(U(1)\) extension, which arises in finite-dimensional quantum mechanics
on a discrete phase space.
By constructing an explicit splitting, it is shown that for \(N=2(2k+1)\),
the group is isomorphic to a semidirect product of \(Sp_N\) and \(\mathbb Z_N^2\).
Moreover, when \(N\) is divisible by \(2^l\ (l\ge2)\), the group has a non-trivial 2-cocycle,
and we provide its explicit form.
Utilizing the splitting, we demonstrate that the corresponding projective Weil representation
can be lifted to a linear representation.
\end{abstract}

%
%
%
%
%

\section{Introduction}

The Wigner function was introduced by E. Wigner
and was used to study the quantum corrections for thermodynamics in 1932 \cite{Wigner}. 
In recent years, its range of applications has extended to
quantum optics and quantum chaos, as well as other fields such as quantum computing.
The concepts derived from the Wigner function have again become a focus of interest
for research in quantum information theory.
For example, the Heisenberg group and its stabilizer group, called the Clifford group, are used
for analyzing quantum error correction.

Among the relevant concepts, the Fano operator plays an important role in Weyl quantization.
We can obtain a quantized Hamiltonian by multiplying a classical Hamiltonian
by the Fano operator and integrating over the phase space.
Thus, the Fano operator can be regarded as a quantization
of a classical phase point, and hence, it is also called a phase point operator.
The Fano operator is obtained from the Fourier transform of the Weyl operator,
which is a group-theoretic consistent lift of classical phase space to the Heisenberg group.
The consistency is related to the automorphism of the Heisenberg group.

The history of the Wigner function on discrete phase space is
relatively young and marked by its application
to a discrete phase space composed of a prime number of lattice points,
formalized by Wootters in 1987 \cite{Wootters},
and to a discrete phase space composed of an odd number of lattice points
corresponding to an integer spin, formalized by Cohendet et al. \cite{Cohendet_1988}. 
However, it was hilighted that its behavior on a discrete phase space,
composed of an even number of lattice points, differs
substantially from that on an odd-lattice phase space.
In 1995, Leonhardt formulated the Wigner function on an even-lattice phase space
corresponding to a half-integral spin but found it necessary to incorporate
a virtual degree of freedom, a so-called ghost variable \cite{Leonhardt_1995,Leonhardt_1996}.
The difficulty was likely due to the different structures of the automorphisms
of the Heisenberg group.

In this paper, we discuss the group structure of automorphisms \(Tp_N\) of the finite Heisenberg group \({\cal E}_N\)
over \(\mathbb Z_N\) with a \(U(1)\) extension.
We can see that \(Tp_N\) is a semidirect product of a symplectic group \(Sp_N\)
whose elements are in a modular class ring \(\mathbb Z_N\) and \(\mathbb Z_N\times\mathbb Z_N\)
when \(N\) is even and not divisible by four.
This is shown through the explicit construction of a splitting \(Sp_N\rightarrow Tp_N\)
using Sunzi's theorem (Chinese remainder theorem).
Using the explicit form, the corresponding projective Weil representation
has a lifting to the Clifford group \(C(N)\). This implies that \(C(N)\) is also a semidirect
product of \(Sp_N\) and \({\cal E}_N\).

\section{Notation and definitions}

\subsection{Residue class ring}
Let \(N\) be a positive natural number and \(\mathbb Z\) be the set of integers.
We denote a quotient ring of \(\mathbb Z\) by \(N\mathbb Z\),
i.e., residue class ring modulo \(N\),
as \(\mathbb Z_N\),
where \(N\mathbb Z\) is ideal in \(\mathbb Z\), composed of
all multiples of \(N\).
An element in \(\mathbb Z_N\) is a set of integers that share the same remainder
when divided by \(N\).
A residue class containing \(\alpha\in\mathbb Z\) is denoted by \([\alpha]_N\),

\begin{equation}
[\alpha]_N=\alpha+N\mathbb Z=\{\alpha+Ni|i\in\mathbb Z\}\in\mathbb Z_N.
\end{equation}

The symbol \([\sbt]_N\) can be regarded as representing a map from \(\mathbb Z\) to \(\mathbb Z_N\),
\begin{equation}
[\sbt]_N: \mathbb Z\longrightarrow\mathbb Z_N,
\end{equation}
\begin{equation}
\alpha\longmapsto[\alpha]_N.
\end{equation}
%
%
We can use the same notation element-wise for matrices,
\begin{equation}
[\sbt]_N:\ M(2,\mathbb Z)\rightarrow M(2,\mathbb Z_N),
\end{equation}
\begin{equation}
A
=
\left(\begin{array}{cc}
\alpha & \beta \\
\gamma & \delta \\
\end{array}\right)
\mapsto[A]_N
=
\left(\begin{array}{cc}
[\alpha]_N & [\beta]_N \\\relax
[\gamma]_N & [\delta]_N \\
\end{array}\right).
\end{equation}

Notation \(
\left[\begin{array}{cc}
\alpha & \beta \\
\gamma & \delta \\
\end{array}\right]_N
\)
is used to abbreviate
\(\displaystyle
\left(\begin{array}{cc}
[\alpha]_N & [\beta]_N \\\relax
[\gamma]_N & [\delta]_N \\
\end{array}\right).
\)
Note that if \(N\) divides an integer \(M\), the map
\begin{equation}
\mathbb Z_M\rightarrow\mathbb Z_N,
\end{equation}
\begin{equation}
[\alpha]_M\mapsto[\alpha]_N,
\end{equation}
is well-defined.
We also denote the map with the same notation \([\sbt]_N\),
\begin{equation}
[[\alpha]_M]_N=[\alpha]_N.
\end{equation}

We have ring isomorphism
\begin{equation}
M(2,\mathbb Z_{N_1})\times M(2,\mathbb Z_{N_2})
\cong
M(2,\mathbb Z_{N_1}\times Z_{N_2})
\end{equation}
by the natural map
\[
\left(
\left(\begin{array}{cc}
[\alpha_1]_{N_1} & [\beta_1]_{N_1} \\\relax
[\gamma_1]_{N_1} & [\delta_1]_{N_1} \\
\end{array}\right),
\left(\begin{array}{cc}
[\alpha_2]_{N_2} & [\beta_2]_{N_2} \\\relax
[\gamma_2]_{N_2} & [\delta_2]_{N_2} \\
\end{array}\right)
\right)
\]
\begin{equation}
\hspace{30mm}
\mapsto
\left(\begin{array}{cc}
([\alpha_1]_{N_1},[\alpha_2]_{N_2}) & ([\beta_1]_{N_1},[\beta_2]_{N_2}) \\
([\gamma_1]_{N_1},[\gamma_2]_{N_2}) & ([\delta_1]_{N_1},[\delta_2]_{N_2}) \\
\end{array}\right),
\end{equation}
and hence, we identify them.

Let \(I_N\) be the set of integers between \(0\) and \(N-1\),
\begin{equation}
I_N=\{0,1,2,\cdots,N-1\}\subset\mathbb Z.
\end{equation}

Each residue class contains a unique element in \(I_N\),
which is denoted as \(\alpha_{(N)}\),
\begin{equation}
\{\alpha_{(N)}\}=[\alpha]_N\cap I_N.
\end{equation}
We call this the basic representative, and
we have \([\alpha]_N=[\alpha_{(N)}]_N\).
The representation \(\sbt_{(N)}\) can be regarded as a map from \(\mathbb Z\) to \(I_N\subset\mathbb Z\),
\begin{equation}
\sbt_{(N)}: \mathbb Z\longrightarrow I_N\subset\mathbb Z,
\end{equation}
\begin{equation}
\alpha\longmapsto\alpha_{(N)}=\alpha-\left[\frac\alpha N\right]N,
\end{equation}
where \(\displaystyle\left[\rule{0pt}{10pt}\ \sbt\ \right]\) is the floor function.
%
%
%
%
We can use the same notation element-wise for matrices,
\begin{equation}
\sbt_{(N)}:\ M(2,\mathbb Z)\rightarrow M(2,I_N)\subset M(2,\mathbb Z),
\end{equation}
\begin{equation}
S
=
\left(\begin{array}{cc}
\alpha & \beta \\
\gamma & \delta \\
\end{array}\right)
\mapsto S_{(N)}
=
\left(\begin{array}{cc}
\alpha_{(N)} & \beta_{(N)} \\\relax
\gamma_{(N)} & \delta_{(N)} \\
\end{array}\right),
\end{equation}
where \(M(2,I_N)\) is a set of \(2\times2\) matrices
composed of elements in \(I_N\), though \(I_N\) is not a ring.
\(
\left(\begin{array}{cc}
\alpha & \beta \\
\gamma & \delta \\
\end{array}\right)_{(N)}
\)
is an abbreviation of 
\(\displaystyle
\left(\begin{array}{cc}
\alpha_{(N)} & \beta_{(N)} \\\relax
\gamma_{(N)} & \delta_{(N)} \\
\end{array}\right).
\)

Note that the map
\begin{equation}
\ I_M\rightarrow I_N,
\end{equation}
\begin{equation}
\alpha_{(M)}\mapsto\alpha_{(N)},
\end{equation}
is well-defined if \(N|M\).
We also denote the map with \(\sbt_{(N)}\),
\begin{equation}
(\alpha_{(M)})_{(N)}=\alpha_{(N)}.
\end{equation}

Here, we introduce the hat symbol defined as
\begin{equation}
\widehat\sbt:\ \mathbb Z\rightarrow\mathbb Z,
\end{equation}
\begin{equation}
\alpha\mapsto\widehat\alpha=\alpha+\left[\frac\alpha2\right]N.
\end{equation}
%
When \(N=2(2k+1)\), we have
\begin{equation}
\widehat\alpha
\equiv
\widehat{\alpha+N}
\equiv
\widehat{\alpha_{(N)}}
\hspace{5mm}\bmod 2N,
\end{equation}
which is used in Sec.4.

\subsection{\label{dblcol}Heisenberg group}
We denote one of the \(N\)-th primitive roots of unity as \(\omega\), defined by
\begin{equation}
\omega=\exp\left(\frac{2\pi\sqrt{-1}}N\right),
\end{equation}
and similarly for \(2N\) as \(\check\omega\),
\begin{equation}
\check\omega=\exp\left(\frac{2\pi\sqrt{-1}}{2N}\right).
\end{equation}
We use \(\omega_N\) when we need to specify \(N\).
Note that \(\omega\) and \(\check\omega\) to the powers of elements in \(\mathbb Z_N\) and \(\mathbb Z_{2N}\), respectively,
are well-defined, but the power of \(\check\omega\) to powers of elements in \(\mathbb Z_N\) is representative-dependent
and not well-defined.

We denote by \({\cal E}_N\), a finite Heisenberg group over \(\mathbb Z_N\)
with a \(U(1)\) extension,
i.e., its elements are in \(U(1)\times\mathbb Z_N\times\mathbb Z_N\),

\begin{equation}
(e^{\sqrt{-1}\theta},[m]_N,[n]_N)\in U(1)\times\mathbb Z_N\times\mathbb Z_N,
\end{equation}
and the composition law is given by
\[
(e^{\sqrt{-1}\theta},[m]_N,[n]_N)\cdot(e^{\sqrt{-1}\theta'},[m']_N,[n']_N)
\]
\begin{equation}
\hspace{30mm}
=
(e^{\sqrt{-1}(\theta+\theta')}\omega^{-[mn']_N},[m+m']_N,[n+n']_N),
\end{equation}
where \(\theta\) is an element of a torus \(T=\mathbb R/(2\pi\mathbb Z)\).
%
In other words, \({\cal E}_N\) is the central extension of \(\mathbb Z_N\times\mathbb Z_N\)
by \(U(1)\) with cocycle \(c\)
\begin{equation}
c(([m]_N,[n]_N),([m']_N,[n']_N))=\omega^{-[mn']_N}.
\end{equation}
%
It is isomorphic to the group formed by the set
\begin{equation}
\{e^{\sqrt{-1}\theta}Q^nP^{-m}|m,n\in\mathbb Z,\theta\in T\}
\end{equation}
with rules
\begin{equation}
Q^N=P^N=I,
\label{qp-definition-cyclicity}
\end{equation}
\begin{equation}
PQ=\omega QP,
\label{qp-definition-commutativity}
\end{equation}
where \(I\) is the identity element and
the \(U(1)\) part commutes with both \(Q\) and \(P\).
The isomorphism is given by
\begin{equation}
(e^{\sqrt{-1}\theta},[m]_N,[n]_N)\mapsto e^{\sqrt{-1}\theta}Q^nP^{-m}.
\end{equation}
In the following, we adopt the notation given on the right side of Eq.(31).

An \(N\)-dimensional irreducible representation, referred to as a Schr\"odinger representation,
is given by
\begin{equation}
P=\sum_{i=0}^{N-1}|i-1\rangle\langle i|,
\hspace{10mm}
Q=\sum_{i=0}^{N-1}|i\rangle\omega^i\langle i|.
\end{equation}
%
The basis is given by
\begin{equation}
\{|i\rangle|i=0,\cdots,N-1\},
\end{equation}
where \(i\) is periodic with modulus \(N\).
It is known that the \(N\)-dimensional representation is unique up to unitary equivalence,
and the unitary matrix is fixed up to the phase.


We consider automorphisms of \(\cal E_N\) leaving the central part \(e^{\sqrt{-1}\theta}\) unchanged,
i.e., center-preserving automorphisms of \(\cal E_N\), which denoted by \(Tp_N\).
%
%
%
%
As we do not change the central part, the element of \(T\in Tp_N\) is
determined by the images of \(Q\) and \(P\).
The general forms of \(T\in Tp_N\) are given for odd \(N\) by
\begin{equation}
T(Q)=Q^{\delta}P^{-\beta}\omega^{-\xi_Q},
\end{equation}
\begin{equation}
T(P)=Q^{-\gamma}P^{\alpha}\omega^{-\xi_P},
\end{equation}
and for even \(N\) by
\[
T(Q)=
\check\omega\ Q^{\delta}P^{-\beta}\omega^{-\xi_Q} \hspace{10mm} {\rm if}\ \delta\ {\rm and}\ \beta: {\rm\ both\ odd},
\]
\begin{equation}
\hspace{19mm}Q^{\delta}P^{-\beta}\omega^{-\xi_Q} \hspace{10mm} {\rm otherwise},
\end{equation}
\[
T(P)=
\check\omega\ Q^{-\gamma}P^{\alpha}\omega^{-\xi_P} \hspace{10mm} {\rm if}\ \gamma\ {\rm and}\ \alpha: {\rm\ both\ odd},
\]
\begin{equation}
\hspace{19mm}Q^{-\gamma}P^{\alpha}\omega^{-\xi_P} \hspace{10mm} {\rm otherwise},
\end{equation}
where
\begin{equation}
S=\left[\begin{array}{cc}\alpha&\beta\\\gamma&\delta\end{array}\right]_N\in Sp_N,
\hspace{10mm}
[\xi_Q,\xi_P]_N\in\mathbb Z_N\times\mathbb Z_N.
\end{equation}

This can be shown using the power formula
\begin{equation}
(P^aQ^b)^c=\check\omega^{abc(c-1)}P^{ac}Q^{bc},
\end{equation}
where \(a,b,c\in\mathbb Z\) are arbitrary integers,
for example,
\begin{equation}
(\check\omega^{-ab(N-1)}P^aQ^b)^N=I,
\label{power-formula-example}
\end{equation}
%

\(Sp_N\) is a group whose elements are in
\(\mathbb Z_N\), and its determinant is \([1]_N\),

\begin{equation}
Sp_N
=
\Set{
\left[\begin{array}{cc}
\alpha & \beta \\
\gamma & \delta \\
\end{array}\right]_N
|
\alpha\delta-\beta\gamma\equiv1
\hspace{5mm}\bmod N
}
\subset M(2,\mathbb Z_N).
\end{equation}

The group \(Sp_N\) is generated from two elements
\begin{equation}
[h_Q]_N
=
\left[\begin{array}{cc}
1 & 1 \\
0 & 1 \\
\end{array}\right]_N,
\hspace{10mm}
[h_P]_N
=
\left[\begin{array}{cc}
1 & 0 \\
1 & 1 \\
\end{array}\right]_N.
\end{equation}

(For example, see Ref.\cite{Watanabe-Hashimoto-Horibe-Hayashi_2019}.)

We also prepare \(Sp_{2N}\) and \(\widetilde{Sp}_{2N}\) as follows for later use:
\begin{equation}
\hspace{-15mm}
Sp_{2N}
=
\Set{
\left[\begin{array}{cc}
\overline\alpha & \overline\beta \\
\overline\gamma & \overline\delta \\
\end{array}\right]_{2N}
|
\overline\alpha\overline\delta-\overline\beta\overline\gamma\equiv1
\hspace{5mm}\bmod 2N
}
\subset M(2,\mathbb Z_{2N}),
\end{equation}
\begin{equation}
\hspace{-15mm}
\widetilde{Sp}_{2N}
=
\Set{
\left[\begin{array}{cc}
\widetilde\alpha & \widetilde\beta \\
\widetilde\gamma & \widetilde\delta \\
\end{array}\right]_{2N}
|
\widetilde\alpha\widetilde\delta-\widetilde\beta\widetilde\gamma\equiv1\hspace{2mm}{\rm or}\hspace{2mm}{1+N}
\hspace{5mm}\bmod 2N
}
\subset M(2,\mathbb Z_{2N}).
\end{equation}
The bar and tilde symbols mean that the corresponding integers are indeterminants of the elements
in \(Sp_{2N}\) and \(\widetilde{Sp}_{2N}\), respectively.
%
%
%

It is easily checked that the map
\begin{equation}
\Pi:Tp_N\longrightarrow Sp_N,
\end{equation}
\begin{equation}
T\mapsto S,
\end{equation}
is a homomorphic projection from \(Tp_N\) to \(Sp_N\).
The kernel \(\Gamma_N\) of \(\Pi\) is composed of elements
\begin{equation}
[\xi]_N\deq[\xi_Q,\xi_P]_N\in\mathbb Z_N\times\mathbb Z_N\deq\Gamma_N,
\end{equation}
%
which is an abelian group.
Hence, we have the following short exact sequence:
\begin{equation}
\{0\}\longrightarrow\Gamma_N\longrightarrow Tp_N\longrightarrow Sp_N\longrightarrow\{1\},
\label{short-exact-sequence-tpn}
\end{equation}
where \(\{0\}\) and \(\{1\}\) are additive and multiplicative trivial groups, respectively, i.e., groups composed of only the identity element.
It is known that if a homomorphism \(\Sigma\) exists from \(Sp_N\) to \(Tp_N\), satisfying
\(\Pi\circ\Sigma={\rm id}|_{Sp_N}\), which is called  `{\it splitting}',
the group \(Tp_N\) is homomorphic to a semidirect product of \(Sp_N\) and \(\Gamma_N\).
Meanwhile, if there is no splitting, the 2-cohomology class becomes non-trivial.
(See Appendix A.)

\section{Odd \(N\) case}
\subsection{Summary of the odd \(N\) case}
To make this paper self-contained, we repeat the result for odd \(N\).
(See Ref.\cite{Appleby_2005}, p.9, Eq.(80) and Ref.\cite{Tolar_2018}, p.6, Eq.(14).)
In this case, we show that \(Tp_N\) is a semidirect product of \(\Gamma_N\) and \(Sp_N\).
\\

\noindent
{\bf Lemma 1}
\vspace{2mm}

Consider the section \(\Sigma\) from \(Sp_N\) to \(Tp_N\)
determined by the following mapping:
\begin{equation}
\Sigma: Sp_N\longrightarrow Tp_N,
\end{equation}
\begin{equation}
[S]_N\longmapsto\Sigma([S]_N)\deq T_{[S]_N},
\end{equation}
where
\begin{equation}
T_{[S]_N}(Q)=\check\omega^{-\delta\beta+\delta\beta N}Q^\delta P^{-\beta},
\label{splitting-odd-Q-prime}
\end{equation}
\begin{equation}
T_{[S]_N}(P)=\check\omega^{-\gamma\alpha+\gamma\alpha N}Q^{-\gamma}P^\alpha.
\label{splitting-odd-P-prime}
\end{equation}
Section \(\Sigma\) is a splitting of the short exact sequence Eq.(\ref{short-exact-sequence-tpn}).
\\

\noindent
{\bf Remark 1}
\vspace{2mm}

The phase part of the power of \(\check\omega\) becomes a power of \(\omega\)
as the factor \(\displaystyle\frac12(1-N)\) is an even integer for odd \(N\).
The matrix
\(\left(\begin{array}{cc}
\delta & -\beta \\
-\gamma & \alpha \\
\end{array}\right)\)
composed of powers of \(Q\) and \(P\) is \(S^{-1}\).
If we define
\begin{equation}
T'_{[S]_N}(Q)=\check\omega^{\alpha\beta-\alpha\beta N}Q^\alpha P^\beta,
\label{splitting-odd-Q}
\end{equation}
\begin{equation}
T'_{[S]_N}(P)=\check\omega^{\gamma\delta-\gamma\delta N}Q^\gamma P^\delta,
\label{splitting-odd-P}
\end{equation}
we have
\begin{equation}
T_{[S]_N}=T'_{[S^{-1}]_N}.
\end{equation}
\ \\

\noindent
{\it (proof of Lemma 1)}

The proof can be shown through calculations.
From Eq.(\ref{power-formula-example}),
it seems
Eqs.(\ref{splitting-odd-Q}) and (\ref{splitting-odd-P})
provide a simple section of the present short exact sequence Eq.(\ref{short-exact-sequence-tpn}).

We can see that the mapping 
satisfies
\begin{equation}
T'_{[S']_N}(T'_{[S]_N}(Q,P))=T'_{[SS']_N}(Q,P),
\end{equation}
comparing the powers of \(Q,P\) and the phases of the two sides,
where
\begin{equation}
T'_{[S]_N}(Q,P)\deq(T'_{[S]_N}(Q),T'_{[S]_N}(P)).
\end{equation}
From Remark 1, we have Lemma 1.

Hence, from Lemma 1 and the general theory of group extension, we have
\begin{equation}
Tp_N\cong Sp_N\ltimes\Gamma_N,
\end{equation}
i.e., \(Tp_N\) is isomorphic to the outer semidirect product of \(\Gamma_N\) and \(Sp_N\).
The action \(\cdot\) of \(Sp_N\) on \(\Gamma_N\) is provided by a matrix product by multiplying by \(\ ^t[S]_N^{-1}\) on the right,
\begin{equation}
[S]_N\cdot[\xi]_N=
[\xi]_N\ ^t[S]_N^{-1}.
\end{equation}

\section{\(N=2(2k+1)\) case}
\subsection{Existence of a splitting}
By Sunzi's theorem (Chinese remainder theorem), \(\mathbb Z_{2(2k+1)}\) can be decomposed
into the direct product of \(\mathbb Z_2\) and \(\mathbb Z_{2k+1}\)
as the factors \(2\)  and \(2k+1\) are coprime:

\begin{equation}
\mathbb Z_{2(2k+1)}\cong\mathbb Z_2\times\mathbb Z_{2k+1}.
\end{equation}

The ring isomorphism is provided by the Sunzi decomposition
\begin{equation}
\phi_N: \mathbb Z_{2(2k+1)}\rightarrow\mathbb Z_2\times\mathbb Z_{2k+1},
\end{equation}
\begin{equation}
[\alpha]_N\mapsto ([\alpha]_2,[\alpha]_{2k+1}).
\end{equation}
(See Appendix B.)

The inverse map of \(\phi_N\) is given by
\begin{equation}
\phi_N^{-1}: \mathbb Z_2\times\mathbb Z_{2k+1}\rightarrow\mathbb Z_{2(2k+1)},
\end{equation}
\begin{equation}
([\alpha]_2,[\beta]_{2k+1})\mapsto[\gamma]_N,
\end{equation}
where
\begin{equation}
\gamma=\alpha_{(2)}\mu_1(2k+1)-\beta_{(2k+1)}\mu_2\cdot2.
\end{equation}
The symbols \(\mu_1\) and \(\mu_2\) are the B\'ezout coefficients satisfying
the B\'ezout identity,
\begin{equation}
\mu_1(2k+1)-\mu_2\cdot2=1,
\end{equation}
for example, \(\mu_1=1,\mu_2=k\).

We have a similar decomposition and composition for \(2N=4(2k+1)\),
\begin{equation}
\mathbb Z_{4(2k+1)}\cong\mathbb Z_4\times\mathbb Z_{2k+1}.
\end{equation}
(See also Appendix C.)



From Eq.(9),
we have
\begin{equation}
M(2,\mathbb Z_{2(2k+1)})\cong M(2,\mathbb Z_2)\times M(2,\mathbb Z_{2k+1}).
\end{equation}

We also use the same notation \(\phi_N\) as in Eq.(61) for the matrices,
which applies the original mapping to each element.
\\

\noindent
{\bf Lemma 2}
\vspace{2mm}

Through the decomposition of the matrix algebra in Eq.(68), we have
\begin{equation}
Sp_{2(2k+1)}\cong Sp_2\times Sp_{2k+1}.
\end{equation}
\ \\

\noindent
{\it (proof)}

The decomposition is easily obtained by restricting the matrix algebra
to the multiplicative group \(Sp_N\).
We consider a map
\begin{equation}
M(2,\mathbb Z_N)\rightarrow M(2,\mathbb Z_2)\times M(2,\mathbb Z_{2k+1}),
\end{equation}
\begin{equation}
\phi_N([S]_N)
\mapsto
([S]_2,[S]_{2k+1}).
\end{equation}
The map is independent of the choice of representative \(S\) and is well-defined.
We can see that the image \(([S]_2,[S]_{2k+1})\) of \([S]_N\in Sp_N\)
is in \(Sp_2\times Sp_{2k+1}\) as the following hold,
\begin{equation}
\phi_N({\rm det}\ [S]_N)
=
({\rm det}\ [S]_2,{\rm det}\ [S]_{2k+1}),
\end{equation}
\begin{equation}
\phi_N([1]_N)
=
([1]_2,[1]_{2k+1}),
\end{equation}
and any element in \(([S]_2,[S]_{2k+1})\)
is an image of \([S]_N\in Sp_N\), and vice versa.
As the map \(\phi_N\) is a ring isomorphism from \(M(2,\mathbb Z_N)\)
to \(M(2,\mathbb Z_2)\times M(2,\mathbb Z_{2k+1})\),
restricting it to \(Sp_N\), we can see the map is a group isomorphism.
Hence, the Lemma is proved.
\\

\noindent
{\bf Remark 2}
\vspace{2mm}

Let \(S\in M(2,\mathbb Z)\) be a matrix satisfying
\begin{equation}
\alpha\delta-\beta\gamma=1,
\end{equation}
i.e., an element in the modular group \(SL(2,\mathbb Z)\).
The map \([\sbt]_N\) is a surjection from \(SL(2,\mathbb Z)\) onto \(Sp(2,N)\);
thus, the representative can be chosen from the modular group.
This is also true for \(([\sbt]_2,[\sbt]_{2k+1})\) and \(Sp_2\times Sp_{2k+1}\).
\\

\noindent
{\bf Lemma 3}
\vspace{2mm}

For \(2N=4(2k+1)\), we have
\begin{equation}
\widetilde{Sp}_{4(2k+1)}\cong\widetilde{Sp}_4\times Sp_{2k+1}.
\end{equation}
\ \\

\noindent
{\it (proof)}

The proof can be provided in a similar way as for Lemma 1, paying attention to
\begin{equation}
[1]_{2N}=([1]_4,[1]_{2k+1}),
\hspace{20mm}
[1+N]_{2N}=([3]_4,[1]_{2k+1}).
\end{equation}
\ \\

We have an isomorphism from \(Sp_2\) to \(\widetilde{Sp}_4\), shown in the following proposition.
\\

\noindent
{\bf Proposition 1}
\vspace{2mm}

The map \(\varphi\),
\begin{equation}
\varphi:\ Sp_2\longrightarrow\widetilde{Sp}_4,
\end{equation}
\begin{equation}
[S]_2=
\left[\begin{array}{cc}
\alpha&\beta\\\relax
\gamma&\delta\\
\end{array}\right]_2
\longmapsto
[\widetilde S]_4=
\left[\begin{array}{cc}
\widetilde\alpha&\widetilde\beta\\\relax
\widetilde\gamma&\widetilde\delta\\
\end{array}\right]_4,
\end{equation}
is an injective homomorphism, where \(\widetilde\alpha,\widetilde\beta,\widetilde\gamma,\widetilde\delta\) are elements in \(I_4\)
defined by
\[
\widetilde S_{(4)}=
\left(\begin{array}{cc}\widetilde\alpha&\widetilde\beta\\\widetilde\gamma&\widetilde\delta\\\end{array}\right)
\deq
\left(\begin{array}{cc}\alpha_{(2)}&\beta_{(2)}\\\gamma_{(2)}&\delta_{(2)}\\\end{array}\right)
+
2\left(\begin{array}{cc}\alpha_{(2)}\gamma_{(2)}&1-\delta_{(2)}\\1-\alpha_{(2)}&\delta_{(2)}\beta_{(2)}\\\end{array}\right)
\]
\begin{equation}
\hspace{7mm}
=S_{(2)}+2(\Delta S).
\end{equation}
\ \\

\noindent
{\it (proof)}

The order of \(Sp_2\) is 6, hence it can be easily checked directly
that the map \(\varphi\) is an isomorphism between \(Sp_2\) and the image of \(\varphi\), i.e., \(Sp_2\cong{\rm Im}\varphi\).
\\

\noindent
{\bf Remark 3}
\vspace{2mm}

Note that \(\Delta S\) can also be written as
\begin{equation}
\Delta S
=
\left(\begin{array}{cc}\alpha_{(2)}-1+\gamma_{(2)}&1-\delta_{(2)}\\1-\alpha_{(2)}&\delta_{(2)}-1+\beta_{(2)}\\\end{array}\right)
\end{equation}
in a linear form.
\\

\noindent
{\it (proof)}

Let
\([S]_N\) be an element in \(Sp_N\),
\begin{equation}
[S]_N=
\left[\begin{array}{cc}
\alpha & \beta \\\relax
\gamma & \delta \\
\end{array}\right]_N,
\hspace{10mm}
\alpha\delta-\beta\delta\equiv1
\hspace{5mm}\bmod N.
\end{equation}
Then,
we have a parity table for integers \(\alpha,\beta, \gamma, \delta\), \\

\begin{center}
\begin{tabular}{cccc}
\hline
 \(\alpha\) & \(\beta\) & \(\gamma\) & \(\delta\) \\
\hline
 \(o\) & \(e\) & \(o\) & \(o\) \\
 \(o\) & \(o\) & \(e\) & \(o\) \\
 \(o\) & \(e\) & \(e\) & \(o\) \\
 \(e\) & \(o\) & \(o\) & \(o\) \\
 \(o\) & \(o\) & \(o\) & \(e\) \\
 \(e\) & \(o\) & \(o\) & \(e\) \\
\hline
\end{tabular}
\end{center}
\ \\
\noindent
where \(e\) and \(o\) imply even and odd numbers, respectively.
Hence, we have the following relations:
\begin{equation}
\alpha_{(2)}\gamma_{(2)}= \alpha_{(2)}+\gamma_{(2)}-1,
\end{equation}
\begin{equation}
\delta_{(2)}\beta_{(2)}= \delta_{(2)}+\beta_{(2)}-1.
\end{equation}
\ \\

\noindent
{\bf Remark 4}
\vspace{2mm}

From Proposition 1 and \([\varphi(\sbt)]_2={\rm id}|_{Sp_2}\), the map \(\varphi\) gives a splitting of the exact sequence,
\begin{equation}
0\longrightarrow\mathbb Z_2^4\xrightarrowi\widetilde{Sp}_4\xrightarrowmd Sp_2\longrightarrow1.
\end{equation}

There are eight splittings.
\\

\noindent
{\it (proof)}

\(Sp_N\) is generated by the two elements
\begin{equation}
[h_Q]_N=
\left[\begin{array}{cc}
1 & 1 \\\relax{}
0 & 1 \\
\end{array}\right]_N,
\hspace{10mm}
[h_P]_N=
\left[\begin{array}{cc}
1 & 0 \\\relax{}
1 & 1 \\
\end{array}\right]_N.
\end{equation}
(See Ref.\cite{Watanabe-Hashimoto-Horibe-Hayashi_2019}.)
Note that an injective homomorphism is determined by its values on the generators.
There are 16 possibilities for the lifted elements by a section for each \([h_Q]_N\) and \([h_P]_N\).
For \(N=2\), restricting the possible elements from the condition that they should have the order \(N\) for a splitting,
only four candidates remain as the splitting images of \([h_Q]_N\) and \([h_P]_N\).
Checking the group structure generated by the \(4^2\) pairs of lifted candidates,
we can obtain eight splittings through direct calculation.
All of them are obtained by the \(A\)-conjugate of the kernel \(\mathbb Z_2^4\),
i.e., 16 kernel elements give eight different splittings.
(See Ref.\cite{Brown}, p.89.)
\\

\noindent
{\bf Corollary 1}
\vspace{2mm}

Composing the maps \(\phi_N, \overline\varphi, \phi_{2N}^{-1}\),
the explicit form of the homomorphism

\begin{equation}
\Phi:\ Sp_{2(2k+1)}\longrightarrow\widetilde{Sp}_{4(2k+1)}
\end{equation}

is given by

\begin{equation}
\left[\begin{array}{cc}
\alpha & \beta \\\relax
\gamma & \delta \\
\end{array}\right]_N
\mapsto
\left[
\left(\begin{array}{cc}
\hat\alpha & \hat\beta \\
\hat\gamma & \hat\delta \\
\end{array}\right)
+
\left(\begin{array}{cc}
\hat\alpha-1+\hat\gamma & 1-\hat\delta \\
1-\hat\alpha & \hat\delta-1+\hat\beta \\
\end{array}\right)N
\right]_{2N},
\end{equation}
where \(\overline\varphi=(\varphi,{\rm id}|_{Sp_{2k+1}})\).
\\

\noindent
{\it (proof)}
 
Composing
\(\phi_N\) and \(\overline\varphi\), we have
\[
\overline\varphi(\phi_N([S]_N))=
\overline\varphi(
\left(
\left[\begin{array}{cc}
\alpha_{(2)} & \beta_{(2)} \\
\gamma_{(2)} & \delta_{(2)} \\
\end{array}\right]_2,
\left[\begin{array}{cc}
\alpha_{(2k+1)} & \beta_{(2k+1)} \\
\gamma_{(2k+1)} & \delta_{(2k+1)} \\
\end{array}\right]_{2k+1}
\right)
)
\]
\[
\hspace{-23mm}
=
\left(
\left[\begin{array}{cc}
\alpha_{(2)} & \beta_{(2)} \\
\gamma_{(2)} & \delta_{(2)} \\
\end{array}\right]_4
+
2\left[\begin{array}{cc}
\alpha_{(2)}-1+\gamma_{(2)} & 1-\delta_{(2)} \\
1-\alpha_{(2)} & \delta_{(2)}-1+\beta_{(2)} \\
\end{array}\right]_4
,
\left[\begin{array}{cc}
\alpha_{(2k+1)} & \beta_{(2k+1)} \\
\gamma_{(2k+1)} & \delta_{(2k+1)} \\
\end{array}\right]_{2k+1}\right)
\]
\[
=
\left(
\left[\begin{array}{cc}
\alpha_{(2)} & \beta_{(2)} \\
\gamma_{(2)} & \delta_{(2)} \\
\end{array}\right]_4
,
\left[\begin{array}{cc}
\alpha_{(2k+1)} & \beta_{(2k+1)} \\
\gamma_{(2k+1)} & \delta_{(2k+1)} \\
\end{array}\right]_{2k+1}\right)
\]
\begin{equation}
+
\left(
2\left[\begin{array}{cc}
\alpha_{(2)}-1+\gamma_{(2)} & 1-\delta_{(2)} \\
1-\alpha_{(2)} & \delta_{(2)}-1+\beta_{(2)} \\
\end{array}\right]_4,
\left[\begin{array}{cc}
0 & 0 \\
0 & 0 \\
\end{array}\right]_{2k+1}
\right).
\end{equation}

Paying attention to
\begin{equation}
[([\alpha_{(2)}]_4,[\alpha_{(2k+1})]_{2k+1)}]_{2N}
=\left[\alpha+\left[\frac\alpha2\right]N\right]_{2N}
=[\hat\alpha]_{2N},
\end{equation}
\begin{equation}
[([2\alpha_{(2)}]_4,[0]_{2k+1})]_{2N}
=[\alpha_{(2)}N]_{2N}
=[\alpha N]_{2N}
=[\hat\alpha N]_{2N},
\end{equation}
etc., element-wise, we have the explicit form in the statement.
(See Appendix D.)
Note that the hat notation is representative-independent in \(\mathbb Z_N\).
(See Eq.(22).)
\\

\noindent
{\bf Proposition 2}
\vspace{2mm}

The following map is a homomorphism from \(\widetilde{Sp}_{2N}\) to \(Tp_N\),
\begin{equation}
\widetilde T: \widetilde{Sp}_{2N}\longrightarrow Tp_N,
\end{equation}
\begin{equation}
[\widetilde S]_{2N}
=
\left[\begin{array}{cc}
\widetilde\alpha & \widetilde\beta \\
\widetilde\gamma & \widetilde\delta \\
\end{array}\right]_{2N}
\longmapsto
\widetilde T_{\widetilde S},
\end{equation}
where
\begin{equation}
\widetilde T_{\widetilde S}(Q)
=
\check\omega^{-\widetilde\delta\widetilde\beta+{\rm stg}(\widetilde S)N}Q^{\widetilde\delta}P^{-\widetilde\beta},
\end{equation}
\begin{equation}
\widetilde T_{\widetilde S}(P)
=
\check\omega^{-\widetilde\gamma\widetilde\alpha+{\rm stg}(\widetilde S)N}Q^{-\widetilde\gamma}P^{\widetilde\alpha},
\end{equation}
and
\({\rm stg}(\widetilde S)\) is a `{\it staggered}' factor of \(\widetilde S\) defined as
\begin{equation}
{\rm stg}(\widetilde S)=\left(\frac{{\rm det}(\widetilde S)-1}N\right)_{(2)}\in I_2.
\end{equation}
\ \\

\noindent
{\it (proof)}

All expressions are independent of the representatives of \(\widetilde S\) and are well-defined.
In this proof, we drop the tilde symbol of \(\widetilde\alpha\) etc., for simplicity,
%
for example,
\begin{equation}
\{\check\omega^{\widetilde\alpha\widetilde\beta+{\rm stg}(\widetilde S)N}P^{\widetilde\alpha}Q^{\widetilde\beta}\}^{\widetilde\gamma}
=
\check\omega^{\alpha\beta\gamma^2+\gamma{\rm stg}(\widetilde S)N}P^{\alpha\gamma}Q^{\beta\gamma}.
\end{equation}

Composing \(\widetilde T_{\widetilde S}\) and \(\widetilde T_{\widetilde S'}\) and using the above formula,
\(Q\) is transformed as
\[
\widetilde T_{\widetilde S}(\widetilde T_{\widetilde S'}(Q))
=\check\omega^{-\delta'\beta'+{\rm stg}(\widetilde S')N}(\widetilde T_{\widetilde S}(Q))^{\delta'}(\widetilde T_{\widetilde S}(P))^{-\beta'}
\]
\[
=\check\omega^{-\delta'\beta'+{\rm stg}(\widetilde S')N}
\{\check\omega^{-\delta\beta+{\rm stg}(\widetilde S)N}
                        Q^\delta P^{-\beta}\}^{\delta'}
\{\check\omega^{-\gamma\alpha+{\rm stg}(\widetilde S)N}
                        Q^{-\gamma}P^\alpha\}^{-\beta'}
\]
\begin{equation}
=
\check\omega^{\kappa^L_Q}Q^{\gamma\beta'+\delta\delta'}P^{-(\alpha\beta'+\beta\delta')},
\end{equation}
where
\begin{equation}
\kappa^L_Q
=-\beta'\delta'-\beta\delta\delta'^2-\alpha\gamma\beta'^2-2\beta\gamma\beta'\delta'
+{\rm stg}(\widetilde S')N
+(\delta'-\beta'){\rm stg}(\widetilde S)N,
\end{equation}
and for \(P\), we have
\[
\widetilde T_{\widetilde S}(\widetilde T_{\widetilde S'}(P))
=\check\omega^{-\gamma'\alpha'+{\rm stg}(\widetilde S')N}(\widetilde T_{\widetilde S}(Q))^{-\gamma'}(\widetilde T_{\widetilde S}(P))^{\alpha'}
\]
\[
=\check\omega^{-\gamma'\alpha'+{\rm stg}(\widetilde S')N}
\{\check\omega^{-\delta\beta+{\rm stg}(\widetilde S)N}
                        Q^\delta P^{-\beta}\}^{-\gamma'}
\{
\check\omega^{-\gamma\alpha+{\rm stg}(\widetilde S)N}
                        Q^{-\gamma}P^\alpha\}^{\alpha'}
\]
\begin{equation}
=
\check\omega^{\kappa^L_P}Q^{-(\gamma\alpha'+\delta\gamma')}P^{\alpha\alpha'+\beta\gamma'},
\end{equation}
where
\begin{equation}
\kappa^L_P
=-\alpha'\gamma'-\beta\delta\gamma'^2-\alpha\gamma\alpha'^2-2\beta\gamma\alpha'\gamma'
+{\rm stg}(\widetilde S')N
+(-\gamma'+\alpha')\:{\rm stg}(\widetilde S)N.
\end{equation}

On the other hand,
\(\widetilde T_{\widetilde S\widetilde S'}\) maps \(Q\) and \(P\) as follows:
\begin{equation}
\widetilde T_{\widetilde S\widetilde S'}(Q)
=
\check\omega^{\kappa^R_Q}Q^{\gamma\beta'+\delta\delta'}P^{-(\alpha\beta'+\beta\delta')},
\end{equation}
where
\begin{equation}
\kappa^R_Q=-(\gamma\beta'+\delta\delta')(\alpha\beta'+\beta\delta')+{\rm stg}(\widetilde S\widetilde S')N,
\end{equation}
and
\begin{equation}
\widetilde T_{\widetilde S\widetilde S'}(P)
=
\check\omega^{\kappa^R_P}Q^{-(\gamma\alpha'+\delta\gamma')}P^{\alpha\alpha'+\beta\gamma'},
\end{equation}
where
\begin{equation}
\kappa^R_P=
-(\gamma\alpha'+\delta\gamma')(\alpha\alpha'+\beta\gamma')
+{\rm stg}(\widetilde S\widetilde S')N.
\end{equation}

The differences of the phase are given by
\begin{equation}
\hspace{-15mm}
\kappa^R_Q-\kappa^L_Q
\hspace{-10mm}
=-\beta'\delta'\:{\rm stg}(\widetilde S)N
+{\rm stg}(\widetilde S\widetilde S')N
-{\rm stg}(\widetilde S')N
-(\delta'-\beta')\:{\rm stg}(\widetilde S)N,
\end{equation}
and
\begin{equation}
\hspace{-15mm}
\kappa^R_P-\kappa^L_P
=-\gamma'\alpha'\:{\rm stg}(\widetilde S)N
+{\rm stg}(\widetilde S\widetilde S')N
-{\rm stg}(\widetilde S')N
-(-\gamma'+\alpha')\:{\rm stg}(\widetilde S)N.
\end{equation}

Paying attention to
\begin{equation}
\beta'\delta'\equiv\beta'+\delta'-1\ {\rm mod}\ 2,
\end{equation}
\begin{equation}
\alpha'\gamma'\equiv \alpha'+\gamma'-1\ {\rm mod}\ 2,
\end{equation}
and a property of the staggered factor
\begin{equation}
{\rm stg}(\widetilde S)
+
{\rm stg}(\widetilde S')
\equiv
{\rm stg}(\widetilde S\widetilde S')\ {\rm mod}\ 2,
\end{equation}
we have
\[
\hspace{-20mm}
\kappa^R_Q-\kappa^L_Q
\equiv
-(\beta'+\delta'-1)\:{\rm stg}(\widetilde S)N
+{\rm stg}(\widetilde S\widetilde S')N
-{\rm stg}(\widetilde S')N
-(\delta'-\beta')\:{\rm stg}(\widetilde S)N
\]
\begin{equation}
\equiv
{\rm stg}(\widetilde S\widetilde S')N
-
{\rm stg}(\widetilde S)N
-
{\rm stg}(\widetilde S')N
\equiv
0
\ {\rm mod}\ 2N
\end{equation}
and
\[
\hspace{-20mm}
\kappa^R_P-\kappa^L_P
\equiv
-(\gamma'+\alpha'-1)\:{\rm stg}(\widetilde S)N
+{\rm stg}(\widetilde S\widetilde S')N
-{\rm stg}(\widetilde S')N
-(-\gamma'+\alpha')\:{\rm stg}(\widetilde S)N
\]
\begin{equation}
\equiv
{\rm stg}(\widetilde S\widetilde S')N
-
{\rm stg}(\widetilde S)N
-
{\rm stg}(\widetilde S')N
\equiv
0
\ {\rm mod}\ 2N,
\end{equation}
which proves the assertion
\begin{equation}
\widetilde T_{\widetilde S}\widetilde T_{\widetilde S'}
=
\widetilde T_{\widetilde S\widetilde S'}.
\end{equation}
\ \\

The homomorphisms are summarized in the following diagram.

\begin{center}
\begin{tikzpicture}
\node (A) at (0,0) {};
\node (B) at (5,0) {\(Sp_{2(2k+1)}\)};
\node (C) at (10,0) {\(Sp_2\times Sp_{2k+1}\)};
\node (D) at (10,3) {\(\widetilde{Sp}_4\times  Sp_{2k+1}\)};
\node (E) at (5,3) {\(\widetilde{Sp}_{4(2k+1)}\)};
\node (F) at (0,3) {\(Tp\)};
\path[draw,->] (B) -- node [below] {Lemma 2} (C);
\path (B) -- node [above] {\(\phi_N\)} (C);
\path[draw,->] (C) -- node [right] {Proposition 1} (D);
\path[] (C) -- node [left] {\(\overline\varphi=(\varphi,{\rm id})\)} (D);
\path[draw,->] (D) -- node [below] {Lemma 3} (E);
\path[] (D) -- node [above] {\(\phi_{2N}^{-1}\)} (E);
\path[draw,->] (E) -- node [below] {Proposition 2} (F);
\path[] (E) -- node [above] {\(\widetilde T\)} (F);
\path[draw,dashed,->] (B) -- node [right] {Corollary 1} (E);
\path[] (B) -- node [left] {\(\Phi\)} (E);
\path[draw,dashed,->] (B) -- node [left] {Corollary 2} (F);
\path[] (B) -- node [right] {\(\Sigma\)} (F);
\end{tikzpicture}
\end{center}

\noindent
{\bf Corollary 2}
\vspace{2mm}

An explicit splitting of the short exact sequence
\begin{equation}
0\rightarrow\mathbb Z_N^2\rightarrow Tp_N\rightarrow Sp_N\rightarrow1
\end{equation}
is given by
\begin{equation}
\Sigma: Sp_N\rightarrow Tp_N,
\end{equation}
\begin{equation}
[S]_N\mapsto\Sigma([S]_N)\deq T_{[S]_N},
\end{equation}
\begin{equation}
T_{[S]_N}(Q)
=
\check\omega^{-\hat\delta\hat\beta+\hat\gamma\hat\alpha N+{\rm stg}(\hat S)N}
Q^{\hat\delta}P^{-\hat\beta},
\end{equation}
\begin{equation}
T_{[S]_N}(P)
=
\check\omega^{-\hat\gamma\hat\alpha+\hat\delta\hat\beta N+{\rm stg}(\hat S)N}
Q^{-\hat\gamma}P^{\hat\alpha},
\end{equation}
where
\begin{equation}
\hat\alpha=\alpha+\left[\frac\alpha2\right]N,
\hspace{10mm}
\hat\beta=\beta+\left[\frac\beta2\right]N,
\end{equation}
\begin{equation}
\hat\gamma=\gamma+\left[\frac\gamma2\right]N,
\hspace{10mm}
\hat\delta=\delta+\left[\frac\delta2\right]N,
\end{equation}
and
\begin{equation}
\hat S=\left(\begin{array}{cc}
\hat\alpha & \hat\beta \\
\hat\gamma & \hat\delta \\
\end{array}\right).
\end{equation}
\ \\

\noindent
{\it (proof)}

From Corollary 1, we have
\[
\Phi([S]_N)
=
\left[
\left(\begin{array}{cc}
\hat\alpha & \hat\beta \\
\hat\gamma & \hat\delta \\
\end{array}\right)
+
\left(\begin{array}{cc}
\hat\alpha-1+\hat\gamma & 1-\hat\delta \\
1-\hat\alpha & \hat\delta-1+\hat\beta \\
\end{array}\right)N
\right]_{2N}
\]
\begin{equation}
\hspace{15mm}
\deq
\widetilde S
=
\left[\begin{array}{cc}
\widetilde\alpha & \widetilde\beta \\
\widetilde\gamma & \widetilde\delta \\
\end{array}\right]_{2N},
\end{equation}
and hence
\begin{equation}
\widetilde\delta\widetilde\beta
\equiv
\hat\delta\hat\beta(1-N)
\ {\rm mod}\ 2N,
\end{equation}
\begin{equation}
\widetilde\gamma\widetilde\alpha
\equiv
\hat\gamma\hat\alpha(1-N)
\ {\rm mod}\ 2N,
\end{equation}
\begin{equation}
{\rm det}(\widetilde S)
=
{\rm det}(\hat S)+(\hat\alpha+\hat\beta+\hat\gamma+\hat\delta)N
\ {\rm mod}\ 2N.
\end{equation}

Using the relations
\begin{equation}
\hat\delta+\hat\beta
\equiv
\hat\delta\hat\beta-1
\ {\rm mod}\ 2,
\hspace{10mm}
\hat\gamma+\hat\alpha
\equiv
\hat\gamma\hat\alpha-1
\ {\rm mod}\ 2,
\end{equation}
we have the explicit form in the statement from Proposition 2.
\\

The results are summarized in Theorem 1 in Sec.6.
\\

\subsection{Lifting of the projective Weil representation}
In this subsection, we consider whether the projective Weil representation \(U([S]_N)\)
induced by the automorphism \(T_{[S]_N}\) in Eqs.(116) and (117) can be lifted to a linear representation.

As
\(Q'=Q_2\otimes Q_{2k+1}\) and \(P'=P_2\otimes P_{2k+1}^{-k}\)
satisfy \(Q'^N=P'^N=I,P'Q'=\omega_NQ'P'\), a unitary matrix \(V\) exists, such that
\begin{equation}
V(Q,P)V^\dagger
=(Q_2\otimes Q_{2k+1},P_2\otimes P_{2k+1}^{-k}),
\end{equation}
where the matrices \(Q_2,P_2\) and \(Q_{2k+1},P_{2k+1}\) are the Schr\"odinger representations
in the dimensions \(2\) and \(2k+1\), respectively.
We have the following proposition as a lifting condition.
\\

\noindent
{\bf Proposition 3}
\vspace{2mm}

If \(VU([S]_N)(Q,P)U^\dagger([S]_N)V^\dagger=VT_{[S]_N}(Q,P)V^\dagger\) is decomposed into automorphisms
in subspaces, such that
\begin{equation}
T_{[S]_N}(Q_2\otimes Q_{2k+1},P_2\otimes P_{2k+1}^{-k})
=(Q_2'\otimes Q_{2k+1}',P_2'\otimes P_{2k+1}'^{-k}),
\end{equation}
\begin{equation}
(Q_2',P_2')=\sigma_2(\lambda_2([S]_N))\;(Q_2,P_2),
\end{equation}
\begin{equation}
(Q_{2k+1}',P_{2k+1}')=\sigma_{2k+1}(\lambda_{2k+1}([S]_N))\;(Q_{2k+1},P_{2k+1}),
\end{equation}
the projective Weil representation \(U([S]_N)\) is liftable,
where
\(\sigma_2\) and \(\sigma_{2k+1}\) are the splittings from \(Sp_2\) to \(Tp_2\) and \(Sp_{2k+1}\) to \(Tp_{2k+1}\), respectively,
and
\(\lambda=(\lambda_2,\lambda_{2k+1})\) is an isomorphism between
\(Sp_N\) and \(Sp_2\times Sp_{2k+1}\).
\\

\noindent
{\it (proof)}

We have a next decomposition into the subspaces for the adjoint action of \(V\),
\begin{equation}
VU([S]_N)(Q,P)U^\dagger([S]_N)V^\dagger
=T_{[S]_N}(Q_2\otimes Q_{2k+1},P_2\otimes P_{2k+1}^{-k}).
\end{equation}
From the assumption, automorphisms
\(\sigma_2(\lambda_2([S]_N))\in Tp_2,
\sigma_{2k+1}(\lambda_{2k+1}([S]_N))\in Tp_{2k+1}\) exist,
which satisfy
\begin{equation}
T_{[S]_N}(Q_2\otimes Q_{2k+1},P_2\otimes P_{2k+1}^{-k})
=(Q_2'\otimes Q_{2k+1}',P_2'\otimes P_{2k+1}'^{-k})
\end{equation}
From the finite-dimensional Stone--von Neumann--Mackey theorem, there exist two and \(2k+1\)-dimensional
unitary matrices \(u_2(\lambda_2([S]_N))\) and \(u_{2k+1}(\lambda_{2k+1}([S]_N))\), respectively,  satisfying
\begin{equation}
\hspace{-10mm}
VU([S]_N)(Q,P)U^\dagger([S]_N)V^\dagger=
\end{equation}
\[
\hspace{-5mm}
(u_2(\lambda_2([S]_N))\otimes u_{2k+1}(\lambda_{2k+1}([S]_N)))
\]
\begin{equation}
\times
(Q_2\otimes Q_{2k+1},P_2\otimes P_{2k+1}^{-k})(u_2(\lambda_2([S]_N))\otimes u_{2k+1}(\lambda_{2k+1}([S]_N)))^\dagger.
\end{equation}
%
It is known that in odd dimensions, \(u_{2k+1}\) is liftable. (See Ref.\cite{Dutta-Prasad_2015}.)
In two dimensions, \(u_2\) can also become liftable through an explicit calculation.
(See Appendix E.)
Denoting the lifted maps as \(\overline u_2\) and \(\overline u_{2k+1}\) and using them to replace
\(u_2\) and \(u_{2k+1}\), we have
\begin{equation}
VU([S]_N)(Q,P)U^\dagger([S]_N)V^\dagger
\end{equation}
\[
=
(\overline u_2(\lambda_2([S]_N)\otimes\overline u_{2k+1}(\lambda_{2k+1}([S]_N)))
\]
\begin{equation}
\hspace{10mm}
\times
V(Q,P)V^\dagger
(\overline u_2(\lambda_2([S]_N)\otimes\overline u_{2k+1}(\lambda_{2k+1}([S]_N)))^\dagger.
\end{equation}
The unitary operator which induces the same automorphism is determined up to phase, and hence,
\begin{equation}
U([S]_N)
\sim
V^\dagger\overline u_2(\lambda_2([S]_N)\otimes\overline u_{2k+1}(\lambda_{2k+1}([S]_N))V.
\end{equation}
The right-hand side is a linear representation, which proves the assertion.
\\

\noindent
{\bf Proposition 4}
\vspace{2mm}

For \([S]_N\in Sp_N\), if we let \(\lambda=(\lambda_2,\lambda_{2k+1})\) be the isomorphism
between \(Sp_N\) and \(Sp_2\times Sp_{2k+1}\),
\begin{equation}
\lambda_2([S]_N)=[S_{(2)}]_2,
\end{equation}
\begin{equation}
\lambda_{2k+1}([S]_N)
=
[S_{2k+1}']_{2k+1}
=
\left[\begin{array}{cc}
\alpha_{(2k+1)} & -k\beta_{(2k+1)} \\
2\gamma_{(2k+1)} & \delta_{(2k+1)} \\
\end{array}\right]_{2k+1},
\end{equation}
and \(\sigma_2, \sigma_{2k+1}\) be
\begin{equation}
\sigma_2=\Sigma{\rm\ in\ Eq.(114)\ with}\ k=0,
\end{equation}
\begin{equation}
\sigma_{2k+1}=\Sigma{\rm\ in\ Eq.(49)},
\end{equation}
then Eq.(127) is satisfied.
The corresponding unitary matrix \(\overline u_2\) is
\begin{equation}
\overline u_2^{++}
\hspace{10mm}
{\rm for\ odd}
\ k,
\end{equation}
\begin{equation}
\overline u_2^{--}
\hspace{10mm}
{\rm for\ even}
\ k
\end{equation}
in Appendix E,
and \(\overline u_{2k+1}\) is determined from 
Eqs.(51), (52)
and \(S_{2k+1}'\).
\\

\noindent
{\it (proof)}

The explicit form of \(VT_{[S]_N}(Q,P)V^\dagger\) is given by
\[
\hspace{-15mm}
VT_{[S]_N}(Q,P)V^\dagger
=
(\omega_{2N}^{-\hat\delta\hat\beta+\hat\gamma\hat\alpha N+{\rm stg}(\hat S)N}
(Q_2^{\delta_{(2)}}
P_2^{-\beta_{(2)}}
\otimes
Q_{2k+1}^{\delta_{(2k+1)}}
P_{2k+1}^{k\beta_{(2k+1)}}),
\]
\begin{equation}
\hspace{20mm}
\omega_{2N}^{-\hat\gamma\hat\alpha+\hat\delta\hat\beta N+{\rm stg}(\hat S)N}
(Q_2^{-\gamma_{(2)}}
P_2^{\alpha_{(2)}}
\otimes
Q_{2k+1}^{-\gamma_{(2k+1)}}
P_{2k+1}^{-k\alpha_{(2k+1)}})).
\end{equation}
Paying attention to
\begin{equation}
\omega_{d_1d_2}^{\alpha\beta}
=
\omega_{d_1}^{\alpha_{(d_1)}\beta_{(d_1)}\mu_1}
\omega_{d_2}^{-\alpha_{(d_2)}\beta_{(d_2)}\mu_2},
\end{equation}
for coprime \(d_1,d_2\), and
\begin{equation}
\omega_4^{\hat\alpha_{(4)}\hat\beta_{(4)}}
=
\omega_4^{\alpha_{(2)}\beta_{(2)}},
\end{equation}
we have
\[
VT_{[S]_N}(Q,P)V^\dagger
\]
\[
\hspace{-20mm}
=
(\omega_{4}^{-\delta_{(2)}\beta_{(2)}\mu_1}
\omega_2^{\gamma_{(2)}\alpha_{(2)}+{\rm stg}(S_{(2)})}
Q_2^{\delta_{(2)}}
P_2^{-\beta_{(2)}}
\otimes
\omega_{2k+1}^{\mu_2\delta_{(2k+1)}\beta_{(2k+1)}}
Q_{2k+1}^{\delta_{(2k+1)}}
P_{2k+1}^{k\beta_{(2k+1)}},
\]
\begin{equation}
\hspace{-15mm}
\omega_{4}^{-\gamma_{(2)}\alpha_{(2)}\mu_1}
\omega_2^{\delta_{(2)}\beta_{(2)}+{\rm stg}(S_{(2)})}
Q_2^{-\gamma_{(2)}}
P_2^{\alpha_{(2)}}
\otimes
\omega_{2k+1}^{\mu_2\gamma_{(2k+1)}\alpha_{(2k+1)}}
Q_{2k+1}^{2k\gamma_{(2k+1)}}
P_{2k+1}^{-k\alpha_{(2k+1)}})).
\end{equation}
In the power of \(Q_{2k+1}\),
\(-\gamma_{(2k+1)}\) is replaced by \(2k\gamma_{(2k+1)}\) for the periodicity of \(Q_{2k+1}\) in modulo \(2k+1\).
For \(S_{(2k+1)}'\), \(t_{2k+1\ [S_{(2k+1)}']_{2k+1}}\), which is \(T_{[S]_N}\) in Eq.(50), is given by
\[
\hspace{-10mm}
(Q_{2k+1}',P_{2k+1}')
\deq
t_{2k+1\ [S_{(2k+1)}']_{2k+1}}(Q_{2k+1},P_{2k+1})
\]
\begin{equation}
\hspace{-10mm}
=
(\omega_{2k+1}^{-k^2\delta_{(2k+1)}\beta_{(2k+1)}}
Q_{2k+1}^{\delta_{(2k+1)}}
P_{2k+1}^{k\beta_{(2k+1)}},
\omega_{2k+1}^{-2k\gamma_{(2k+1)}\alpha_{(2k+1)}}
Q_{2k+1}^{\delta_{(2k+1)}}
P_{2k+1}^{k\beta_{(2k+1)}}),
\end{equation}
so
\(P_{2k+1}'^{-k}\) is 
\begin{equation}
(t_{2k+1\ [S_{(2k+1)}']_{2k+1}}(P))^{-k}
=
\omega_{2k+1}^{-k^2\gamma_{(2k+1)}\alpha_{(2k+1)}}
Q_{2k+1}^{2k\gamma_{(2k+1)}}
P_{2k+1}^{-k\alpha_{(2k+1)}}.
\end{equation}
Using these equations, we have
\[
VT_{[S]_N}(Q,P)V^\dagger
\]
\begin{equation}
\hspace{-10mm}
=
(
Q_2'
\otimes
\omega_{2k+1}^{(\mu_2+k^2)\delta_{(2k+1)}\beta_{(2k+1)}}
Q_{2k+1}',
\hspace{3mm}
P_2'
\otimes
\omega_{2k+1}^{(\mu_2+k^2)\gamma_{(2k+1)}\alpha_{(2k+1)}}
P_{2k+1}'^{-k}),
\end{equation}
where
\begin{equation}
(Q_2',P_2')=t_2^s(Q_2,P_2),
\end{equation}
and the sign \(s\) of \(t_2\) is given by
\begin{equation}
s=++\hspace{10mm}{\rm for\ odd}\ k,
\end{equation}
\begin{equation}
s=--\hspace{10mm}{\rm for\ even}\ k.
\end{equation}
(See Appendix E.)
It is easy to see that the power of \(\omega_{2k+1}\) satisfies
\begin{equation}
\mu_2+k^2\equiv0
\hspace{5mm}\bmod 2k+1
\end{equation}
for both odd and even \(k\).
Therefore,
\(VT_{[S]_N}(Q,P)V^\dagger\) is expressed in \(Q_2',P_2'\) and \(Q_{2k+1}',P_{2k+1}'\) as
\begin{equation}
VT_{[S]_N}(Q,P)V^\dagger
=
(Q_2'\otimes Q_{2k+1}',P_2'\otimes(P_{2k+1}')^{-k}),
\end{equation}
which proves the assertion.
\\

The results are summarized in Theorem 2 in Sec.6.

\section{\(N=2^l(2k+1),l\ge2\) case}

Let \([h_Q]_N\) and \([h_J]_N\) be
\begin{equation}
[h_Q]_N
=
\left[\begin{array}{cc}
1 & 1 \\
0 & 1 \\
\end{array}\right]_N,
\hspace{10mm}
[h_J]_N
=
\left[\begin{array}{cc}
0 & 1 \\
-1 & 0 \\
\end{array}\right]_N,
\end{equation}
\([h_Q]_N,[h_J]_N\in Sp_N\).
We drop the notation \([\sbt]_N\) for simplicity in this section.
We have
\begin{equation}
h_Q^N=I,
\hspace{10mm}
h_J^2h_Q=h_Qh_J^2,
\hspace{10mm}
h_J^2h_Q=h_Qh_J^2
=
\left[\begin{array}{cc}
-1 & -1 \\
 0 & -1 \\
\end{array}\right]_N.
\end{equation}
The general forms of \(T_{h_Q}\) and \(T_{h_J}\) are given by
\begin{equation}
T_{h_Q}(Q)=\omega_{2N}\omega_N^aQP^{-1},
\hspace{10mm}
T_{h_Q}(P)=\omega_N^bP,
\end{equation}
and
\begin{equation}
T_{h_J}(Q)=\omega_N^cP^{-1},
\hspace{10mm}
T_{h_Q}(P)=\omega_N^dQ,
\end{equation}
respectively. Repeating \(T_{h_Q}\) \(k\) times, we have
\begin{equation}
T_{h_Q}^k(Q)
=
\omega_{2N}^k\omega_N^{ka}\omega_N^{-b\frac12k(k-1)}QP^{-k}.
\end{equation}
As \(T_{h_Q}^N=I\), setting \(k=N\),
\begin{equation}
\omega_N^{-b\frac12N(N-1)}=-1
\end{equation}
is obtained.
The values of \(\displaystyle\frac12N(N-1)\) are classified modulo \(4\) as
\begin{equation}
\frac12N(N-1)\equiv2l
\hspace{10mm}
{\rm for}
\hspace{5mm}
N=4l,
\end{equation}
\begin{equation}
\frac12N(N-1)\equiv0
\hspace{10mm}
{\rm for}
\hspace{5mm}
N=4l+1,
\end{equation}
\begin{equation}
\frac12N(N-1)\equiv2l+1
\hspace{10mm}
{\rm for}
\hspace{5mm}
N=4l+2,
\end{equation}
\begin{equation}
\frac12N(N-1)\equiv0
\hspace{10mm}
{\rm for}
\hspace{5mm}
N=4l+3.
\end{equation}
Hence,
\begin{equation}
\omega^{-2bl}=-1
\end{equation}
for \(N=2^l(2k+1),l\ge2\).
On the other hand, from
\begin{equation}
T_{h_Q}T_{h_J}^2(P)=T_{h_J}^2T_{h_Q}(P),
\end{equation}
we have
\begin{equation}
\omega_N^{-2b}=1,
\end{equation}
which leads to a contradiction with \(\omega^{-2bl}=-1\).
Hence, no value of \(b\) satisfies the above relation.
Note that when \(N=2(2k+1)\), \(b\equiv2k+1
\hspace{1mm}\bmod N\).
This means there is no splitting
from \(Sp_N\) to \(Tp_N\) if \(N\) is divisible by four.
Hence, we have the next proposition.
\\

\noindent
{\bf Proposition 5}
\vspace{2mm}

\(Tp_N\) is not a semidirect product of \(\mathbb Z_N\times\mathbb Z_N\) and \(Sp_N\)
when \(N\) is divisible by four.
\\

Let \(T_{\pm,\pm}\) be four elements in \(Tp_N\), such that
\begin{equation}
T_{\pm,\pm}(Q,P)=(\pm Q,\pm P).
\end{equation}
When we consider a simple section
\begin{equation}
Sp_N\rightarrow Tp_N,
\end{equation}
\begin{equation}
[S]_N\mapsto T_{[S_{(N)}]_{2N}},
\end{equation}
the corresponding 2-cocycle \(C\) defined as
\begin{equation}
T_{[S_1]_N}T_{[S_2]_N}=T_{[S_1S_2]_N}C([S_1]_N,[S_2]_N)
\end{equation}
is easily calculated.
The result is given by Proposition 6.
\\

\noindent
{\bf Proposition 6}
\vspace{2mm}

The 2-cocycle is given by
\begin{equation}
C([S_1]_N,[S_2]_N)=T_{c_1,c_2},
\end{equation}
\begin{equation}
\left(\begin{array}{cc}
\alpha' & \beta' \\
\gamma' & \delta' \\
\end{array}\right)
\deq
(S_1)_{(N)}(S_2)_{(N)},
\end{equation}

\begin{equation}
c_1
=
(-1)^{
\llbracket\delta',\beta'\rrbracket
+
\llbracket\alpha',\delta'\rrbracket
+
\llbracket\beta',\delta'\rrbracket
},
\end{equation}
\begin{equation}
c_2
=
(-1)^{
\llbracket\gamma',\alpha'\rrbracket
+
\llbracket\alpha',\delta'\rrbracket
+
\llbracket\beta',\delta'\rrbracket
},
\end{equation}
\begin{equation}
\llbracket \alpha,\beta\rrbracket=\langle\alpha\rangle\beta+\alpha\langle\beta\rangle,
\end{equation}
where
\(\langle\alpha\rangle\) is 
defined as
\(\langle\alpha\rangle\deq\displaystyle\frac{\alpha_{(2N)}-\alpha_{(N)}}N\).
\\

For this simple section, it is not easy to see that the 2-cocycle is a boundary for \(N=2(2k+1)\).

\section{Summary}

We investigated the group structure of the automorphisms \(Tp_N\)
of the finite Heisenberg group \({\cal E}_N\) over \(\mathbb Z_N\)
with a \(U(1)\) extension. The results are summarized in Theorem 1.
\\

\noindent
{\bf Theorem 1}
\vspace{2mm}

If \(N=2k+1\) or \(N=2(2k+1)\),
\begin{equation}
Tp_N\cong Sp_N\ltimes\Gamma_N.
\end{equation}
The action of \(Sp_N\) on \(\Gamma_N\) is given by multiplying
by \(\ ^tS^{-1}\) on the right,
\begin{equation}
[S]_N\cdot[\xi]_N=[\xi]_N\ ^tS^{-1}.
\end{equation}
If \(N=2^l(2k+1)\ (l\ge2)\), then there is no splitting, and hence,
\(Tp_N\) has no trivial 2-cohomology class.
\\

For the projective Weil representation of the homomorphic section \(T_{[S]_N}\),
we have Theorem 2.
\\

\noindent
{\bf Theorem 2}
\vspace{2mm}

If \(N=2k+1\) or \(N=2(2k+1)\), there is a choice of phase to render
the projective representation \(U([S]_N)\) induced by \(T([S]_N)\) linear.
Hence, the projective Weil representation \(U([S]_N)\) of \(Sp_N\) is liftable to be linear.
\\

The stabilizer group of \({\cal E}_N\) in \(N\)-dimensional unitary group \(U(N)\)
is called the Clifford group \(C(N)\).
Since
\begin{equation}
1\rightarrow{\cal E}_N\rightarrow C(N)\rightarrow Sp_N\rightarrow1
\end{equation}
is a short exact sequence,
Theorem 2 implies that there exists a splitting from \(Sp_N\) to \(C(N)\).
Hence, \(C(N)\) is a semidirect product of \(Sp_N\) and \({\cal E}_N\),
\begin{equation}
C(N)=Sp_N\ltimes{\cal E}_N.
\end{equation}
\ \\

\ack
The authors are greatful to Prof. J. Tolar for sending us useful
literatures and suggestive nformation at the early stage of this work.

\vspace{10mm}

\noindent
{\bf Note added}
\vspace{2mm}

When we were preparing the manuscript, we found a work by M. Korbel\'ar and J. T. Tolar,
Ref.\cite{Korbelar-Tolar_2023},
in which they obtained the same result as Theorem 1 through a detailed analysis
of group presentation.

\newpage

\appendix

\section{General theory of group extension}

Let \(\{0\}\) and \(\{1\}\) be an additive group and a multiplicative group, respectively, each composed of a single identity element,
and let \(A\) be an additive group and \(E\), \(G\) be groups.
Letting the notation \(\longrightarrow\) mean a group homomorphism,
the short exact sequence


\begin{equation}
\{0\}\longrightarrow A\xrightarrowi E\xrightarrowpi G\longrightarrow\{1\}
\end{equation}


\noindent
is called a group extension.
Refer to Ref.\cite{Brown} Chapter IV, Low Dimensional Cohomology and Group Extension, pp. 86--106.
Group \(E\) is called an extended group of \(G\) by \(A\).
The {\it 'exact'} means the kernel of the right-hand side homomorphism is the same
as the image of left-hand side homomorphism, e.g.,
\begin{equation}
{\rm Im}\:i={\rm Ker}\:\pi.
\end{equation}
We can see that \(\pi\) is surjective and \(i\) is injective.

A set-theoretic section \(s\) from \(G\) to \(E\) is defined by a map
\begin{equation}
s:G\longrightarrow E
\end{equation}
satisfying
\begin{equation}
\pi\circ s={\rm id},
\end{equation}
where \({\rm `id'}\) is an identity map on \(G\).

It is known that
if there exists a homomorphic section, which we call a {\it splitting}, 
the extended group \(E\) is isomorphic to a semidirect product of \(G\) and \(A\),
\begin{equation}
E\cong A\rtimes G.
\end{equation}
See \cite{Brown} pp.86--89, especially Proposition (2.1) on p.87. \\

The left action \(\cdot\) of \(G\) on \(A\) is defined by
\begin{equation}
i(g\cdot a)=\tilde gi(a)\tilde g^{-1},
\end{equation}
where \(a\in A, g\in G\) and \(\tilde g\in\pi^{-1}(g)\).
This definition is independent of the choice of \(\tilde g\)
when \(A\) is an abelian group.
The explicit group product law is given by
\begin{equation}
(a,g)(a',g')=(a+g\cdot a',gg'),
\end{equation}
and the isomorphism from \(A\rtimes G\) to \(E\) is
\begin{equation}
(a,g)\mapsto i(a)\sigma(g).
\end{equation}
If there is no homomorphic section, the group product law is specified by
a 2-cocycle depending on the section.
The group structure of \(E\) itself is fixed by the second cohomology class
\begin{equation}
H^2(G,A).
\end{equation}
(See Ref.\cite{Brown} pp. 91--94 for more details.)

\newpage

\section{Sunzi theorem, Sunzi representation, Sunzi representative}

Sunzi's theorem is given as follows.
\\

\noindent
{\bf Sunzi's theorem}

Let \(m_1,m_2,\cdots,m_k\) be \(k\) coprime integers,
\(m_i\perp m_j\ (i\ne j)\).
Then, there exists a unique integer \(x\) modulo \(m_1m_2\cdots m_k\),
such that
\begin{equation}
x\equiv a_i
\hspace{5mm}\bmod m_i,
\hspace{10mm}
i=1,2,\cdots,k.
\end{equation}
\ \\

This theorem is also known as the Chinese remainder theorem.
Generally, let \(R\) be a unital ring and
\(I_i\ (1\le i\le k)\) be coprime two-sided ideals.
Then, we have a ring isomorphism
\begin{equation}
R/I\cong\prod_{i=1}^kR/I_i.
\hspace{10mm}
I=\bigcap_{i=1}^kI_i.
\end{equation}

In the simple case of \(k=2\), we have
\begin{equation}
\mathbb Z/(d_1d_2\mathbb Z)
\cong
\mathbb Z/(d_1\mathbb Z)\times\mathbb Z/(d_2\mathbb Z),
\end{equation}
namely
\begin{equation}
\mathbb Z_{d_1d_2}\cong\mathbb Z_{d_1}\times\mathbb Z_{d_2},
\end{equation}
for residue class rings when \(d_1\perp d_2\).
We denote the isomorphism as \(\phi_{d_1,d_2}\),
\begin{equation}
\phi_{d_1,d_2}: \mathbb Z_{d_1d_2}\rightarrow Z/(d_1\mathbb Z)\times\mathbb Z/(d_2\mathbb Z),
\end{equation}
\begin{equation}
[\alpha]_{d_1d_2}\mapsto([\alpha]_{d_1},[\alpha]_{d_2}).
\end{equation}
We denote it simply as \(\phi_N\) with \(N=d_1d_2\)
if \(d_1\) and \(d_2\) are obvious in the context.

In this case, we can find two integers \(\nu_1\) and \(\mu_2\)
satisfying
\begin{equation}
\mu_1d_2-\mu_2d_1=1,
\end{equation}
which are called as the B\'ezout coefficients.
There are two basic solutions
\((\mu_1^{(+)},\mu_2^{(+)})\)
and
\((\mu_1^{(-)},\mu_2^{(-)})\)
for the B\'ezout coefficients satisfying
\begin{equation}
|\mu_1|\le|d_1|,
\hspace{10mm}
|\mu_2|\le|d_2|,
\end{equation}
\begin{equation}
\mu_1^{(+)}>0,\ \mu_2^{(+)}>0,
\hspace{10mm}
\mu_1^{(-)}<0,\ \mu_2^{(-)}<0.
\end{equation}

Let \(([\alpha]_{d_1},[\alpha]_{d_2})\) be the Sunzi decomposition of \([\alpha]_{N}\),
\begin{equation}
\phi_N([\alpha]_N)=([\alpha]_{d_1},[\alpha]_{d_2}).
\end{equation}
Using the B\'ezout coefficients, the inverse map of \(\phi_N\) can be
written as follows:
\begin{equation}
[\gamma]_{N}=\phi_N^{-1}(([\alpha]_{d_1},[\beta]_{d_2}))
=[\alpha_{(d_1)}\mu_1d_2-\beta_{(d_2)}\mu_2d_1]_N,
\end{equation}
where \(\alpha_{(d_1)}\) and \(\beta_{(d_2)}\) are
elements in \(I_{d_1}\) and \(I_{d_2}\), respectively.
We call this inverse map the Sunzi composition.

\newpage

\section{B\'ezout coefficients for \(N=2(2k+1)\) and \(2N\)}
\vspace{2mm}

In the \(N=2(2k+1)\ (k\ge1)\) case, the B\'ezout coefficients satisfying
\begin{equation}
\mu_1(2k+1)-\mu_2\cdot2=1
\end{equation}
are given by
\begin{equation}
\mu_1=1,
\hspace{10mm}
\mu_2=k.
\end{equation}
In the \(2N=4(2k+1)\) case, the B\'ezout coefficients satisfying
\begin{equation}
\mu_1(2k+1)-\mu_2\cdot4=1
\end{equation}
depend on whether \(k\) is odd or even.
If \(k\) is odd, i.e., \(k=2l-1\ (l\ge1)\),
the B\'ezout equation is
\begin{equation}
\mu_1(4l-1)-\mu_2\cdot4=1
\end{equation}
\begin{equation}
\therefore\ )\ \ 
\mu_1=-1,
\hspace{10mm}
\mu_2=-l.
\end{equation}
If \(k\) is even, i.e., \(k=2l\ (l\ge2)\),
the B\'ezout equation becomes
\begin{equation}
\mu_1(4l+1)-\mu_2\cdot4=1
\end{equation}
\begin{equation}
\therefore\ )\ \ 
\mu_1=1,
\hspace{10mm}
\mu_2=l,
\end{equation}

\newpage

\section{Eq.(89) and Eq.(90)}

First, we observe
\begin{equation}
\alpha_{(4)}\equiv\alpha_{(2)}+\left[\frac\alpha2\right]2
\hspace{5mm}\bmod4.
\end{equation}
Note that \(N=2(2k+1)\) in this case.
\\

For Eq.(89), we have
\begin{equation}
[([\alpha_{(2)}]_4,[\alpha_{(2k+1)}]_{2k+1})]_{2N}
=
[\alpha_{(2)}\mu_1(2k+1)-\alpha_{(2k+1)}\mu_2\cdot4]_{2N}.
\end{equation}
When \(k=2l-1\), \(\mu_1=-1\) and \(\mu_2=-l\),
\[
\alpha_{(2)}\mu_1(2k+1)-\alpha_{(2k+1)}\mu_2\cdot4
=
-\alpha_{(2)}(4l-1)+\alpha_{(4l-1)}\cdot4l
\]
\[
\equiv
-(\alpha_{(4)}-\left[\frac\alpha2\right]2)(4l-1)+\alpha_{(4l-1)}\cdot4l
\]
\[
=
(-\alpha_{(4)}(2k+1)+\alpha_{(2k+1)}\cdot2k)
+\left[\frac\alpha2\right]2(2k+1)
\]
\begin{equation}
\equiv
\alpha+\left[\frac\alpha2\right]N
=
\hat\alpha
\hspace{5mm}\bmod2N.
\end{equation}

\noindent
When \(k=2l\), \(\mu_1=1\) and \(\mu_2=l\), then
\[
\alpha_{(2)}\mu_1(2k+1)-\alpha_{(2k+1)}\mu_2\cdot4
=
\alpha_{(2)}(4l+1)-\alpha_{(4l+1)}\cdot4l
\]
\[
\equiv
(\alpha_{(4)}-\left[\frac\alpha2\right]2)(4l+1)-\alpha_{(4l+1)}\cdot4l
\]
\[
=
(\alpha_{(4)}(2k+1)-\alpha_{(2k+1)}\cdot2k)
-\left[\frac\alpha2\right]2(2k+1)
\hspace{5mm}\bmod2N.
\]
\begin{equation}
\equiv
\alpha-\left[\frac\alpha2\right]N
\equiv
\alpha+\left[\frac\alpha2\right]N
\equiv
\hat\alpha
\hspace{5mm}\bmod2N.
\end{equation}

\begin{equation}
\therefore\ )\ \ 
[([\alpha_{(2)}]_4,[\alpha_{(2k+1)}]_{2k+1})]_{2N}
=
[\hat\alpha]_{2N}.
\end{equation}

For Eq.(90),
\[
[([2\alpha_{(2)}]_4,[0]_{2k+1})]_{2N}
=
[2\alpha_{(2)}\mu_1(2k+1)-0\cdot\mu_2\cdot4]_{2N}
\]
\begin{equation}
=
[\alpha_{(2)}\mu_1\cdot2(2k+1)]_{2N}
=
[\alpha_{(2)}\mu_1N]_{2N}
\equiv
[\alpha_{(2)}N]_{2N},
\end{equation}
as \(\mu_1=-1\) or \(+1\) for odd or even \(k\), respectively.
On the other hand,
\begin{equation}
[\hat\alpha N]_{2N}
=
\left[\left(\alpha+\left[\frac\alpha2\right]N\right)N\right]_{2N}
=
\left[\alpha N\right]_{2N}
=
\left[\alpha_{(2)}N\right]_{2N}.
\end{equation}

\begin{equation}
\therefore\ )\ \ 
[([2\alpha_{(2)}]_4,[0]_{2k+1})]_{2N}
=
[\hat\alpha N]_{2N}.
\end{equation}

\newpage

\section{Weil representation of \(Sp_2\)}

There are four splittings from \(Sp_2\) to \(Tp_2\).
The splittings are fixed by the values of two generators
\begin{equation}
[h_Q]_2=\left[\begin{array}{cc}
 1 &  1  \\
 0 &  1 \\
\end{array}\right]_2,
\hspace{10mm}
[h_P]_2=\left[\begin{array}{cc}
 1 &  0  \\
 1 &  1 \\
\end{array}\right]_2.
\end{equation}
For the four splittings \(t_2^{\tau_Q\tau_P}\ (\tau_Q=\pm,\tau_P=\pm)\),
the values are given by
\begin{equation}
t^{\tau_Q\tau_P}_{2\ [h_Q]_2}(Q,P)=(\tau_Q\sqrt{-1}QP,-P),
\end{equation}
\begin{equation}
t^{\tau_Q\tau_P}_{2\ [h_P]_2}(Q,P)=(-Q,\tau_P\sqrt{-1}QP),
\end{equation}
where the superscripts \(\tau_Q\) and \(\tau_P\) are
signs of the \(QP\) parts in the upper and lower equations, respectively.

It can be checked that each \(t_2^{\pm\pm}\) can be generated from the adjoint action of
the next \(\overline u_2^{\pm\pm}\),

\begin{equation}
\hspace{-23mm}
t_2^{++}:
\hspace{1mm}
\overline u_2^{++}([h_Q]_2)=\frac1{\sqrt2}\left(\begin{array}{cc}
 1 &  i  \\
-i & -1 \\
\end{array}\right),
\hspace{1mm}
\overline u_2^{++}([h_P]_2)=\frac1{\sqrt2}\left(\begin{array}{cc}
 0   & -1-i \\
-1+i &  0   \\
\end{array}\right),
\end{equation}
\begin{equation}
\hspace{-23mm}
t_2^{--}:
\hspace{1mm}
\overline u_2^{--}([h_Q]_2)=\frac1{\sqrt2}\left(\begin{array}{cc}
-1 &  i  \\
-i &  1 \\
\end{array}\right),
\hspace{1mm}
\overline u_2^{--}([h_P]_2)=\frac1{\sqrt2}\left(\begin{array}{cc}
 0   &  1+i \\
 1-i &  0   \\
\end{array}\right),
\end{equation}
\begin{equation}
\hspace{-23mm}
t_2^{+-}:
\hspace{1mm}
\overline u_2^{+-}([h_Q]_2)=\frac1{\sqrt2}\left(\begin{array}{cc}
 1 &  i  \\
-i & -1 \\
\end{array}\right),
\hspace{1mm}
\overline u_2^{+-}([h_P]_2)=\frac1{\sqrt2}\left(\begin{array}{cc}
 0   &  1-i \\
 1+i &  0   \\
\end{array}\right),
\end{equation}
\begin{equation}
\hspace{-23mm}
t_2^{-+}:
\hspace{1mm}
\overline u_2^{-+}([h_Q]_2)=\frac1{\sqrt2}\left(\begin{array}{cc}
-1 &  i  \\
-i &  1 \\
\end{array}\right),
\hspace{1mm}
\overline u_2^{-+}([h_P]_2)=\frac1{\sqrt2}\left(\begin{array}{cc}
 0   & -1-i \\
-1+i &  0   \\
\end{array}\right).
\end{equation}
These unitary matrices determine the projective Weil representation of \(Sp_2\)
and they are shown to form a linear representation.
This indicates that the projective Weil representation is liftable for \(N=2\).

\newpage

\section*{References}

\end{document}